\documentclass[12pt,aastex]{article}

\newcommand{\nat}{{\it Nature}}
\newcommand{\mnras}{{\it MNRAS}}

\usepackage{multicol}
\usepackage{authblk}
\usepackage{color}
\usepackage{natbib} 
\usepackage{graphicx}
\usepackage{amssymb} 
\usepackage{verbatim} 
\usepackage{tabularx}
\usepackage{wrapfig}
\usepackage{sectsty}
\usepackage{enumitem}
\setlist[itemize]{leftmargin=*}
\def\aap{A\&A}
\def\apj{ApJ}
\def\apjl{ApJL}
\sectionfont{\fontsize{14}{16}\selectfont}

\title{\vspace{-20mm}\Large\bf Astro 2020 White Paper: \\
Realizing the Unique Potential of ALMA to Probe the Gas Reservoir of Planet Formation\\ 
{\normalsize {\bf Thematic Area:  Star and Planet Formation}}
\vspace{-2mm}}
\normalsize
\author[1]{Ilse Cleeves}
\author[2]{Ryan Loomis}
\author[3]{Richard Teague}
\author[3]{Ke Zhang}
\author[3]{Edwin Bergin}
\author[4]{Karin \"Oberg}
\author[2]{Crystal Brogan}
\author[2]{Todd Hunter}
\author[5]{Yuri Aikawa}
\author[4]{Sean Andrews}
\author[6]{Jaehan Bae}
\author[4]{Jennifer Bergner}
\author[7]{Kevin Flaherty}
\author[8]{Viviana Guzman}
\author[4]{Jane Huang}
\author[9]{Michiel Hogerheijde}
\author[10]{Shih-Ping Lai}
\author[11]{Laura P\'erez}
\author[4]{Charlie Qi}
\author[12]{Luca Ricci}
\author[13]{Colette Salyk}
\author[14]{Kamber Schwarz}
\author[15]{Jonathan Williams}
\author[4]{David Wilner}
\author[2]{Al Wootten}

\affil[1]{\small Principal Author, University of Virginia, USA, lic3f@virginia.edu} 
\affil[2]{\small National Radio Astronomy Observatory, USA}
\affil[3]{\small University of Michigan, USA}
\affil[4]{\small Harvard University, USA}
\affil[5]{\small University of Tokyo, Japan}
\affil[6]{\small Carnegie Institution of Washington, DTM, USA}
\affil[7]{\small Williams College, USA}
\affil[8]{\small Pontificia Universidad Catolica de Chile, Chile}
\affil[9]{\small Leiden University, Netherlands}
\affil[10]{National Tsing Hua University, Taiwan}
\affil[11]{Universidad de Chile, Santiago, Chile}
\affil[12]{\small California State University, USA}
\affil[13]{\small Vassar College, USA}
\affil[14]{\small University of Arizona, USA}
\affil[15]{\small University of Hawaii, USA}

\date{}
\begin{document}
\maketitle
\vspace{-15mm}

\thispagestyle{empty}

\begin{comment}
\begin{abstract}
\normalsize 
Understanding the origin of the astonishing diversity of exoplanets is a key question for the coming decades.  ALMA has revolutionized our view of the dust emission from protoplanetary disks, demonstrating the prevalence of ring and spiral structures that are likely sculpted by young planets in formation.  To detect kinematic signatures of these protoplanets and to probe the chemistry of their gas accretion reservoir will require the imaging of molecular spectral line emission at high angular and spectral resolution.  However, the current sensitivity of ALMA limits these important spectral studies to only the nearest protoplanetary disks.  Although some promising results are emerging, including the identification of the snowlines of a few key molecules and the first attempt at detecting a protoplanet's spiral wake, it is not yet possible to search for these important signatures in a population of disks in diverse environments and ages. 
Harnessing the tremendous power of (sub)mm observations to pinpoint and characterize the chemistry of planets in formation will require a major increase of ALMA’s spectral sensitivity ($5-10\times$), increase in instantaneous bandwidth ($2\times$) at high spectral resolution, and improved angular resolution ($2\times$) in the 2030 era.
\end{abstract}
%\end{titlingpage}
\end{comment}

\medskip

\noindent{\bf Abstract:} Understanding the origin of the astonishing diversity of exoplanets is a key question for the coming decades.  ALMA has revolutionized our view of the dust emission from protoplanetary disks, demonstrating the prevalence of ring and spiral structures that are likely sculpted by young planets in formation.  To detect kinematic signatures of these protoplanets and to probe the chemistry of their gas accretion reservoir will require the imaging of molecular spectral line emission at high angular and spectral resolution.  However, the current sensitivity of ALMA limits these important spectral studies to only the nearest protoplanetary disks.  Although some promising results are emerging, including the identification of the snowlines of a few key molecules and the first attempt at detecting a protoplanet's spiral wake, it is not yet possible to search for these important signatures in a diverse population of protoplanetary disks. 
Harnessing the tremendous power of (sub)mm observations to pinpoint and characterize the chemistry of planets in formation will require a major increase of ALMA’s spectral sensitivity ($5-10\times$), increase in bandwidth ($2\times$) at high spectral resolution, and improved angular resolution ($2\times$) in the 2030 era.

\clearpage
\setcounter{page}{1}

\section{Introduction}
\vspace{-3mm}
\begin{wrapfigure}{r}{0.46\textwidth}
\vspace{-1.3cm} 
\hspace{-0.1cm}
\includegraphics[width=0.99\linewidth]{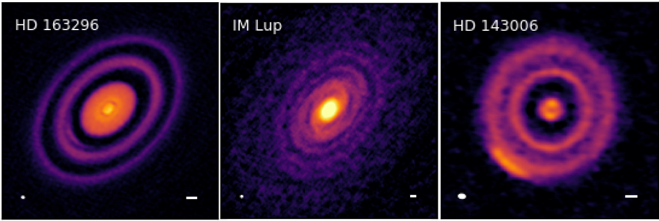}
\begin{center}
\vspace{-0.8cm}
\centering
\caption{\em Exquisite ALMA dust continuum images at 35~mas (5~au) angular resolution from DSHARP \citep{Andrews2018}.}
\label{dsharp}
\vspace{-1.2cm}
\end{center}
\end{wrapfigure}

Today $>3900$ exoplanets are confirmed, and this number will continue to grow with current ({\em  TESS} and {\em GAIA}) and future missions ({\em JWST, PLATO, WFIRST}, etc). Even given current observational biases, the diversity of exoplanetary system architectures is astonishing. ALMA is currently leading a revolution in our understanding of the {\em origins} of this diversity, allowing us for the first time to peer deep into protoplanetary disks and capture images of planet formation in action. (Sub)millimeter dust continuum observations reveal the evolution of disk midplane solids as protoplanets form, as exemplified by a recent ALMA Large Program (DSHARP; Fig.~\ref{dsharp}) of 20 disks revealing numerous dark/bright rings, spiral structures, and azimuthal asymmetries (with typical size scales of 5 -- 10~au) that are generally thought to be sculpted by the presence of hidden planets in their infancy. However, the dust can only tell a fraction of the story: it is the gas that traces $99\%$ of a protoplanetary disk's mass, encodes all of the kinematic information, and reveals the chemical reservoir for planet formation. 
With ALMA, we are just now beginning to unlock the unique diagnostic potential of gas-phase spectroscopic observations and link the physical {\em and} chemical properties of protoplanetary disks with their forming planets.

As we enter a new era when the characterization of exoplanetary atmospheres becomes routine, ALMA shows promise to be a transformative instrument in connecting exoplanets with the story of their origins. Achieving this potential, however, will require both spatially and spectrally resolving key diagnostic line emission at relevant physical scales (such data are inherently $\sim 2$ orders of magnitude less sensitive than the continuum). Moreover, such studies must cover a representative sample of disks that span a range of evolutionary states, disk morphologies, and environments. Here the current limitations of ALMA become apparent. Presently, to achieve $\sim10-15$~au resolution for spectroscopic study of {\em only five} targets requires a 130 hr ALMA Large Program (PI: {\"O}berg). These disks reside at distances of $\sim140$~pc, but in order to study the closest disks in a massive star forming environment (e.g., Orion), we must reach out to $\sim400$~pc. Improving spectral surface brightness sensitivity and simultaneous bandwidth to observe more diagnostic lines at once will therefore be critical for comprehensive spectral studies of  protoplanetary disks in the coming decades.

Here we present key science drivers for spectroscopic study of protoplanetary systems in the (sub)mm regime, highlighting the present state of the art and areas where deficiencies in current capabilities motivate the significant upgrades outlined in the ALMA 2030 Development Roadmap \citep{Carpenter2019}. In particular, we show that a 5-$10\times$ increase in spectral sensitivity coupled with an increase in spectral agility and bandwidth will both dramatically improve our capability to directly detect protoplanets {\em and} massively expand the sample size of surveys investigating the chemical environment in which exoplanets form.

\vspace{-5mm}
\section{Kinematic Detection of Planets in Formation}
\vspace{-3mm}
To directly confront planet formation theories, we must find planets during their formation, while still embedded in the disk. Previously, there have been two main approaches to this goal. The first exploits the high angular resolution of extreme adaptive optics (XAO; e.g.\ Gemini Planet Imager and SPHERE/VLT) to try to detect thermal emission from the young planet, or H$\alpha$ emission from accretion \citep{Wagner2018}. However, searches in nearby disks with significant mm dust continuum substructure have resulted in many upper limits, suggesting that protoplanets are generally much cooler, or accrete significantly less vigorously than predicted. The second approach is the detection of circumplanetary disks (CPD) at (sub)mm wavelengths. Though \citet{Zhu2018} predict ALMA could detect CPDs down to 0.03~lunar masses, CPD emission has not yet been detected \citep[e.g.,][]{Andrews2018}. 

The spectroscopic imaging power of ALMA has led to a new approach for planet detection through searches for gas kinematic perturbations due to the gravitational influence of embedded protoplanets \citep{Perez2018, Pinte2018, Teague2018a}.
Embedded 
\begin{wrapfigure}{r}{0.5\textwidth}
\vspace{-0.5cm} % vspace hspace order matters
\hspace{-0.25cm}
    \includegraphics[width=0.96\linewidth]{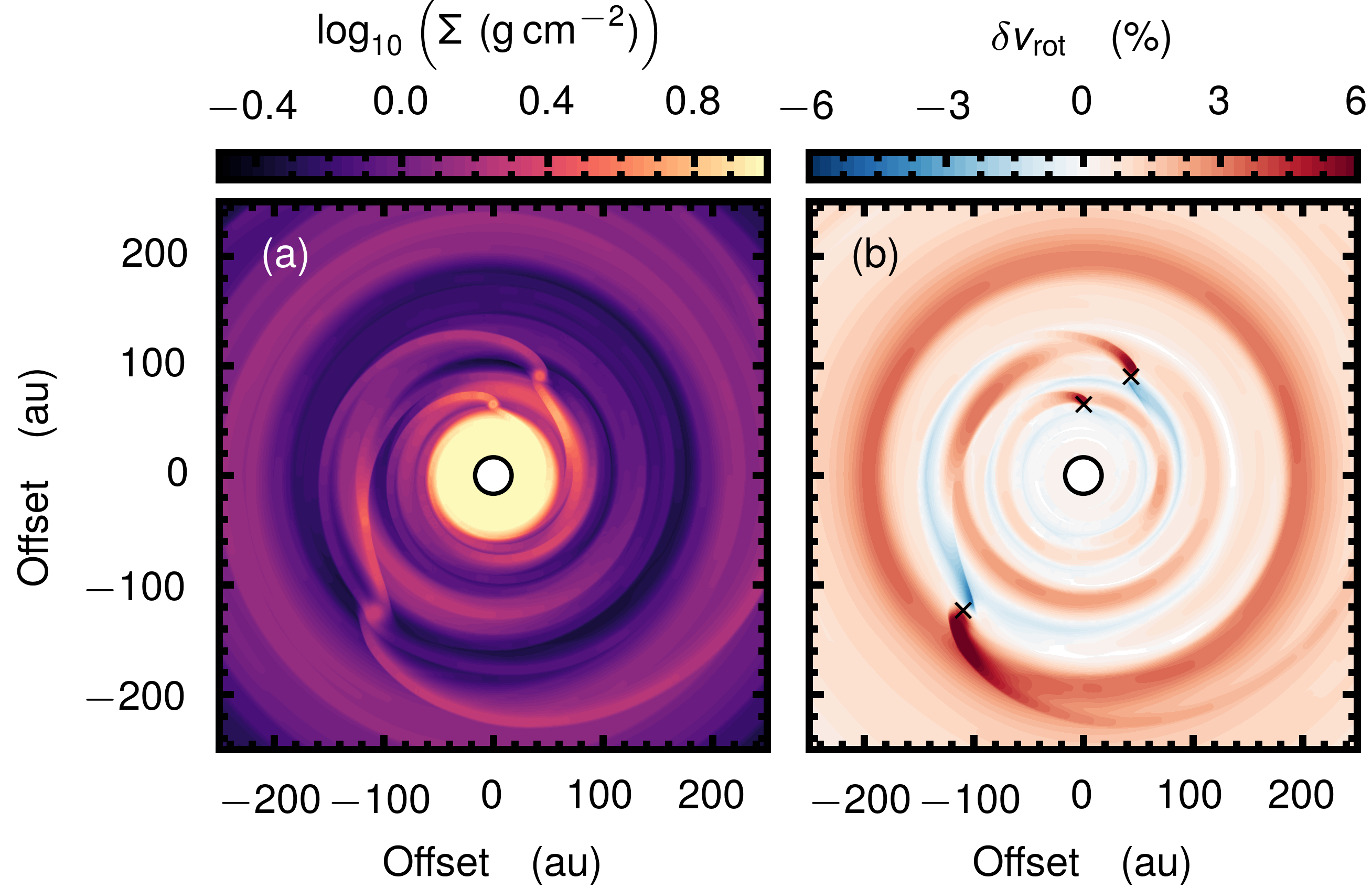}
\begin{center}
\vspace{-0.75cm}
\centering
\caption{{\it Hydrodynamic model of three 1~$M_{\rm Jup}$ planets showing gas (a) density and (b) kinematics revealing distinct wake patterns \citep{Teague2018a}.}}
    \label{fig:HD163296hydro}
\vspace{-1.2cm}
\end{center}
\end{wrapfigure}
planets  drive spiral wakes, resulting in local density enhancements and changes in the gas velocity due to the gas pressure gradient (Fig.~\ref{fig:HD163296hydro}). These effects result in two clear observables. First, the planet clears some material along its orbit, creating a gas deficit. 
Second, density variations perturb the radial pressure gradient and the rotational velocity of the gas \citep{Kanagawa2015}, an effect which has already been identified in a 
handful of sources \citep{Teague2018a,Teague2018c}. 

Though intriguing, {\em a more definitive method will be directly imaging the spiral pattern of the wake} (Fig.~\ref{fig:HD163296hydro}b). Identifying  wakes provides two significant advantages. First, since detection will not be limited to the inner disk regions where the mm grains reside,  protoplanet searches can extend to the entire {\em gas} disk. At larger radial separations, studies in the NIR (e.g. {\em JWST} or ELTs) will also be feasible as contamination from the stellar PSF
\begin{wrapfigure}{r}{0.53\textwidth}
\vspace{-1.2cm} 
\hspace{-0.3cm}
    \includegraphics[width=1\linewidth]{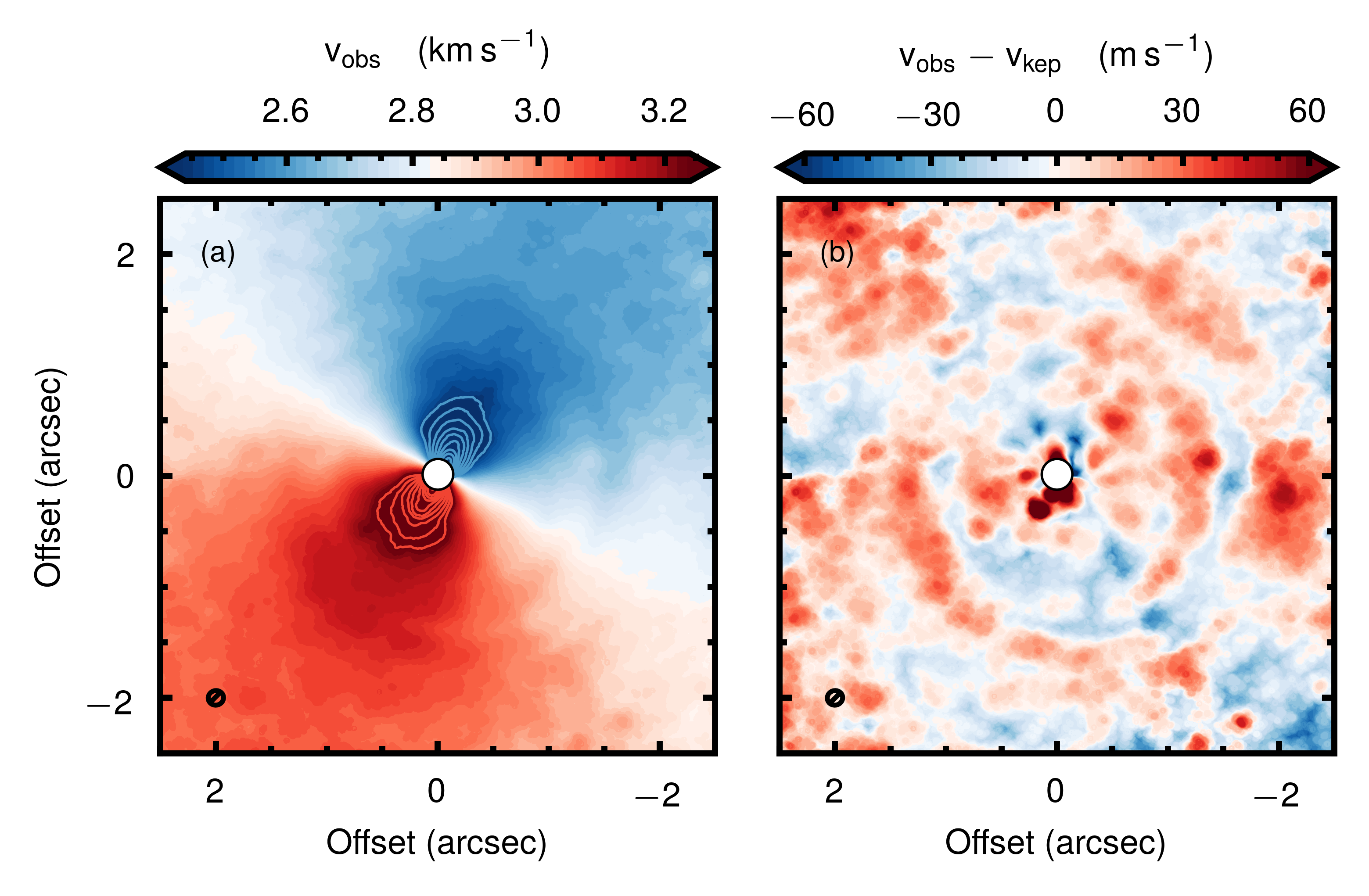}
\begin{center}
\vspace{-0.96cm}
\centering
\caption{{\it (a) $^{12}$CO(3-2) line-of-sight velocity around TW~Hya with 8~au resolution \citep{Huang2018a}.
        (b) Residual after removing bulk Keplerian motion, revealing azimuthal structure, hinting at planet-driven features.}}
    \label{fig:TWHya}
\vspace{-1.2cm}
\end{center}
\end{wrapfigure}
  is reduced. Second, wake signatures are typically {\em larger} spatially than CPDs, 
making them accessible at lower spatial resolution.

However, ALMA currently lacks the sensitivity required to resolve  spatial scales comparable to the ring/gap structures in the dust continuum ($\sim 5$~au) for any but the most nearby disks. Fig.~\ref{fig:TWHya} demonstrates the current state of the art in high angular resolution kinematic studies, with 6.6~hr on-source time towards the nearest disk TW\,Hya (d=60 pc) in $^{12}$CO(3-2), and 8~au resolution. Hints of azimuthal structures are observed, albeit amid significant noise.
Confirmation will require significantly more integration time even toward nearby TW\,Hya so that more optically thin tracers can be used. Exploiting the true power of this technique in a sample of protoplanetary disks (unavoidably at larger distances) will require both high angular resolution (at least $2\times$) to achieve the requisite 5~au resolution and significantly higher sensitivity to overcome the commensurate decrease in surface brightness sensitivity.

\vspace{-5.0mm}
\section{The Chemical Environment of Forming Planets}
\vspace{-3.5mm}

\begin{figure}[h!]
\centering
\begin{minipage}{0.67\linewidth}
\includegraphics[width=0.98\linewidth]{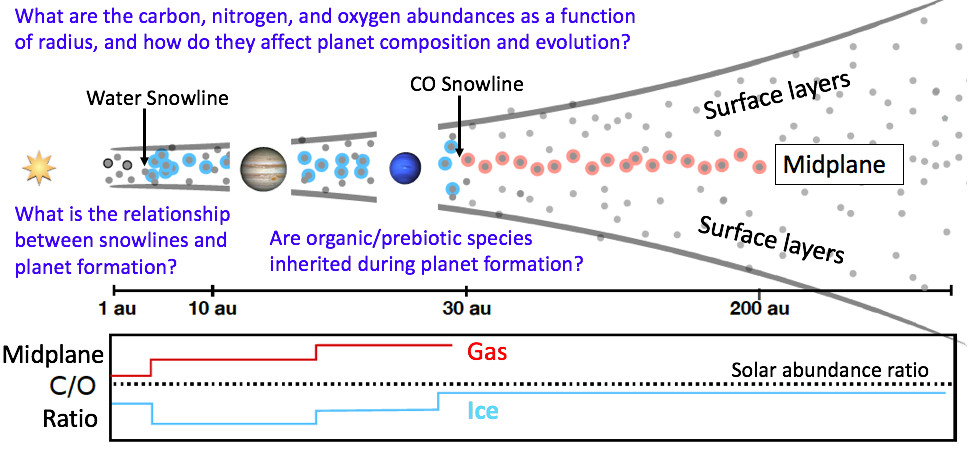}
\end{minipage}
\begin{minipage}{0.32\linewidth} 
\caption{{\it {\bf Top} Cartoon of the radial distribution of key disk components. {\bf Bottom} Midplane C/O ratio prediction compared to Solar for gas and ice. The C/O ratio changes radially due to the freeze out of species like H$_2$O, CO$_2$, and CO \citep{Oberg2011}.}}\label{fig:c2o}
\end{minipage}\vspace{-3.0mm}
\end{figure}

The chemistry and physics of planet formation are intimately linked  (Fig.~\ref{fig:c2o}), and we are just beginning to scratch the surface of this connection.  With ALMA we can now directly observe snowlines where volatiles freeze out of the gas phase, and we can probe the indirect effects of physical evolution on chemistry.
Even with observations limited to a handful of the most nearby protoplanetary disks, it is rapidly becoming clear that their chemistry is {\em actively evolving}. Some of the strongest evidence for these deviations from a simple inherited interstellar chemistry comes from synergistic ALMA and {\em Herschel} observations, showing respectively that both CO and water vapor are strongly depleted in disk surfaces compared to  interstellar abundances \citep{hogerheijde2011,miotello2017,du2017}.

\vspace{-3mm}

These tantalizing results suggest that the evolution of the disk chemical environment may play an important role in setting the range of planetary compositions, but many of the most crucial observations of gas are prohibitively expensive and thus currently limited in scope and sample size. {\em We still do not know what the most common disk compositions are, and therefore we do not know what the most probable exoplanet compositions are likely to be.} As exoplanet atmospheric characterization capabilities rapidly improve \citep[e.g.,][]{madhusudhan2018}, such information will be critical in designing programs for follow-up atmospheric characterization of confirmed exoplanets from missions such as {\em TESS}.

Below we describe three key science questions for uncovering the chemical environment of planet formation. First, it is increasingly clear that the observed composition of disk surface layers is inconsistent with that of earlier interstellar stages \citep[e.g.,][]{cleeves2018iau}. It is therefore crucial to trace the evolution of disk chemistry in a statistically significant sample of sources across a wide range of physical environments and ages. Second, investigations of the interface between disk surface layers and the icy grains in the planet-forming midplane will be critical to interpreting the impact of gaseous chemical evolution on planetary inheritance. Here, direct and indirect ALMA observations of snowlines will be highly complementary with upcoming infrared studies of the disk ices and inner disk gas. Finally, emission from complex organic species in disks is inherently weak, but offers a powerful tool to constrain the interstellar inheritance of prebiotic material. An increase in surface brightness sensitivity at sub(mm) wavelengths would be transformative for each of these goals, and expanded instantaneous bandwidths would allow many to be achieved simultaneously.

\smallskip
%\vspace{-0.7mm}

\noindent {\bf What is the range of possible disk compositions, and which are common?}
The leading explanation for the aforementioned differences between the gas-phase carbon, oxygen, nitrogen, and sulfur abundances and interstellar values is that the volatiles are being sequestered into ice-coated grains that grow into larger pebbles or even bodies such as comets or planetesimals. This process preferentially removes oxygen (in the form of water) from the observable surface layers of the disk \citep{bergin2016,cleeves2018iau}, which enhances the C/O ratio in the gas \citep[][Fig.~\ref{fig:c2o}]{Oberg2011}. Under high C/O conditions, abundant hydrocarbons such as C$_2$H will form \citep{du2015}, suggesting that observations of these hydrocarbons may be useful as a proxy for tracing disk chemical evolution. For example, the older ($\sim 8$ Myr) disk TW Hya requires a C/O ratio $\gtrsim 1$ to reproduce the brightness of the observed C$_2$H lines \citep{bergin2016}, while the younger IM\,Lup ($\sim0.5$ Myr) disk, only requires C/O~$\sim 0.8$. Similarly, observations of optically thin N-bearing species such H$^{13}$CN can be used to constrain the disk nitrogen content \citep{cleeves2018}.

Furthermore, with upcoming observations anticipated from {\em JWST}, we will be able to search for the ``missing'' ices at the same radii that ALMA probes the gas using broadband ice absorption features \citep{aikawa2012}, and also test for radial transport of icy-coated dust grains into the terrestrial planet forming region by investigating volatile chemistry in the inner disk with {\em JWST} MIRI. For example, if the evolving grains transport extensive amounts of water into the inner disk, we should be able to see an inner gas-phase water enhancement, which would enrich the atmospheres of forming giant planets, potentially explaining close in gas giant exoplanets with water rich atmospheres \citep[e.g.,][]{pinhas2018}.

However, we are still in the regime of small number statistics, limited in our ability to detect key species sensitive to C/N/O like C$_2$H and isotopologues of HCN and CO toward a large sample of disks ($\sim$ a few hundred). ALMA surveys have had relatively few detections of the CO isotopologues compared to models with interstellar abundances \citep{ansdell2016}. By improving ALMA's spectral line sensitivity, we have the potential to unlock {\em in a statistical way} what are the most common compositions planets can inherit from their disks.

\smallskip
\noindent {\bf How do snowlines mediate the chemical and physical disk evolution?} 
 The freeze-out of different volatiles (H$_2$O, CO$_2$, and CO) as ice onto dust grains may dramatically improve the ability of grains to coagulate into larger bodies \citep{Ros2013,Banzatti2015} and also shifts the balance of ice- versus gas-phase carbon, oxygen, nitrogen, etc., directly impacting the resulting initial chemical composition that a forming planet may inherit \citep[see Fig.~\ref{fig:c2o} and][]{Oberg2011}. However, complicating this picture, if dust grains have grown to sufficiently large sizes, they may start to ``blur'' the specific snowline locations as the grains drift inward \citep{Piso2015,Piso2016}. Therefore {\em direct measurements} of snowline locations are critical for identifying the locations of these threshold regions (Fig.~\ref{fig:c2o}). 
 
The midplane CO snowline around sun-like stars is expected to occur between $10-40$~au, readily accessible with ALMA. Peering through highly optically-thick surface layers down to the midplane, however, requires the use of weakly emitting, optically-thin isotopologues as tracers. $^{13}$C$^{18}$O has emerged as a promising diagnostic, successfully employed by \citet{Zhang2017} to unambiguously identify the mid-plane CO snowline at 21~au in TW\,Hya (d=60~pc) with ALMA. Similar studies for a larger sample of T\,Tauri disks is not feasible with the current sensitivity, however; imaging $^{13}$C$^{18}$O in a single disk at the distance of Taurus (d=140~pc) would require $\sim 30$ hr on-source integration time. A 5-10$\times$ increase in spectral sensitivity would allow surveys of minimum-mass solar nebula type disks (60~M$_\oplus$ of solids plus 0.01~$M_\odot$ of H/He) across a number of local star-forming regions.

Directly accessing the H$_2$O midplane snowline is more challenging for ALMA because of 
its compact radial distribution \citep[within $1-5$~au;][]{Zhang2013,Blevins2016}, 
and a lack of optimal transitions. However, several weak warm/hot ($E_{U}\sim$ 100s to 1000s of K) transitions of H$_2$O, and H$_2^{18}$O in the (sub)mm offer hope for detecting, and even resolving the distribution of water at larger radii along its snow-surface interface. A tentative detection of H$_2$O and H$_2^{18}$O at 321-322 GHz has been reported in a disk 120~pc away \citep{Carr2018}. Only with a more sensitive ALMA can we push these studies forward, connecting water observations at larger radii with observations closer to the star from facilities such as {\em JWST} 
to provide a cohesive picture of water chemistry across a large sample.

\smallskip
\vspace{-0.5mm}

\noindent {\bf What is our interstellar organic inheritance?} 
It is currently unclear whether the molecular inventory of disks, particularly the midplane, is set by interstellar inheritance or an active disk chemistry.
During the early prestellar phase, a rich chemistry has already begun, including abundant water and  organics \citep{jimenezserra2016,caselli2010}. Models suggest that some material, including water and organics, can be preserved in disks \citep[e.g.,][]{visser2009,cleeves2014wat,cleeves2016org,drozdovskaya2018}. 

Although organic molecules are widely observed at earlier stages of star formation, low inherent gas-phase column densities makes their detection challenging in protoplanetary disks.  The deep integrations required, however, pay off with optically thin emission, which allows the gas-phase organic properties to be observed throughout the vertical extent of the disk, including closer to the midplane if non-thermal desorption is efficient. 
Moreover, these species' closely spaced lines  enable key disk physical properties like temperatures and densities to be constrained, fundamentally anchoring physical models. 

ALMA has provided the first detections of ``complex'' organics like CH$_3$CN, CH$_3$OH, and HCOOH toward nearby protoplanetary disks \citep{Oberg_2015, Walsh_2016, Favre_2018}. Observations of CH$_3$CN show the strong potential of organics as unambiguous tracers of excitation conditions \citep{Loomis_2018,Bergner_2018}. 
Even these observations, however, are limited by prohibitively large integration times and lower resolutions, restricting our understanding at planet forming spatial scales.
A 5-10$\times$ better spectral line sensitivity would enable organics to be used as a powerful probe of disk inheritance {\em and} physical/kinematic structure (\S2) across a larger sample of disks. Larger instantaneous bandwidths ($\geq2\times$) would allow more diagnostic transitions to be observed at once, enabling 
all the key science goals described here to be simultaneously achievable.

\vspace{-5mm}
\section{Recommendations}
\vspace{-3mm}

ALMA is leading a revolution in our understanding of planet formation. 
Nonetheless, the current limited spectral surface brightness sensitivity of ALMA restricts the study of the crucial {\em gas} component of planet formation to a handful of the most nearby objects in a non-representative sample of environments. In order to harness the tremendous power of (sub)mm observations to pinpoint and chemically characterize planets in formation requires a 5-10$\times$ improvement of ALMA's spectral sensitivity and increased bandwidth ($\geq2\times$) at high spectral resolution for simultaneous observation of diagnostic lines in the 2030 era. These goals can be realized with a combination of increased collecting area, improved receivers, and increasing the bandwidth, efficiency, and data rates of the ALMA signal processing system.

\newpage

\begin{multicols}{2}
\setlength{\bibsep}{-0.0pt}

\vspace{-0.3cm}
\bibliographystyle{myapj}
\end{multicols}

\end{document}